%% file: main.tex
\documentclass{llncs}
\usepackage{latexsym}
\usepackage{amsfonts}
\usepackage{amsmath}
\usepackage{epsfig}
\usepackage{graphics}
\usepackage{verbatim}

\newtheorem{clm}{Claim}
\newtheorem{thm}{Theorem}
\newtheorem{defn}{Definition}
\newtheorem{cor}[thm]{Corollary}
\newtheorem{prop}[thm]{Proposition}

\newcommand{\E}{{\mathbb E}}

\newcommand{\R}{{\mathbb R}}

\newcommand{\hn} {\hat{n}}
\newcommand{\hN} {\hat{N}}

\newcommand{\trans} {\mathsf{T}}
\newcommand{\var} {\mbox{var}}
\newcommand{\prob} {{\bf P}}

\newcommand{\xvec}{{\mathbf{x}}}
\newcommand{\yvec}{{\mathbf{y}}}
\newcommand{\zvec}{{\mathbf{z}}}
\newcommand{\Myz}{{M(\yvec,\zvec)}}

\begin{document}
\title{Competition-Induced Preferential Attachment}
\author{
N.~Berger\inst{1} \and 
C.~Borgs\inst{1} \and
J.~T.~Chayes\inst{1} \and
R.~M.~D'Souza\inst{1} \and
R.~D.~Kleinberg\inst{2}}
\institute{Microsoft Research, One Microsoft Way, Redmond WA 98052, USA
\and
M.I.T. CSAIL, 77 Massachusetts Ave, Cambridge MA 02139, USA.  Supported
by a Fannie and John Hertz Foundation Fellowship.}
\maketitle \def\thepage {}

\input{abstract}

\input{intro1}
\input{intro2}

\input{equivalence}
\input{convexp}

\input{powerlaw}
\bibliographystyle{plain}

\input{main.bbl}
\pagebreak
\appendix
\input{appdx1}
\input{appdx2}
\input{appdx3}

\input{appdx4}
\end{document}

%% file: abstract.tex











\begin{abstract}
Models based on preferential attachment have had much success in
reproducing the power law degree distributions which seem ubiquitous in
both natural and engineered systems.  Here, rather than assuming
preferential attachment, we give an explanation of how it can arise
from a more basic underlying mechanism of competition between
opposing forces.

We introduce a family of one-dimensional geometric growth models,
constructed iteratively by locally optimizing
the tradeoffs between two competing metrics.  This family
admits an equivalent description
as a graph process
with no reference to the
underlying geometry.
Moreover, the resulting graph process is shown to be preferential
attachment with an upper cutoff.
We rigorously determine the degree distribution
for the family of random graph
models, showing that it obeys a power law up to a finite threshold and
decays exponentially above this threshold.

We also introduce and rigorously analyze a generalized version of
our graph process, with two natural parameters, one corresponding
to the cutoff and the other a ``fertility'' parameter.  Limiting
cases of this process include the standard Barab\'asi-Albert
preferential attachment model and the uniform attachment model. In
the general case, we prove that the process has a power law degree
distribution up to a cutoff, and establish monotonicity of the
power as a function of the two parameters.

\end{abstract}

%% file: intro1.tex


\section{Introduction}
\subsection{Network Growth Models}

There is currently tremendous interest in understanding the mathematical
structure of networks -- especially as we discover how  pervasive
network structures are in natural and engineered systems.  Much recent
theoretical work has been motivated by measurements of real-world
networks, indicating they have certain ``scale-free'' properties, such
as a power-law distribution of degree sequences.
For the Internet
graph, in particular, both the graph of
routers  and the graph of autonomous systems (AS) seem to
obey  power
laws \cite{Falout3,GovinTangmu2000}.  However, these
observed power laws hold
only for a limited range of degrees, presumably due to physical constraints
and the finite size of the Internet.

Many random network growth models have been proposed which
give rise to power law degree distributions.  Most of these models
rely on a small number of basic mechanisms, mainly
preferential attachment%
\footnote{As Aldous \cite{Aldous-SCN}
points out, proportional attachment may be
a more appropriate name, stressing the linear dependence of
the attractiveness on the
degree.}
\cite{Price,BarabasiAlbert99} or
copying \cite{KRRSTU}, extending ideas known for many
years \cite{Polya,Simon,Zipf,Yule} to a network context.
Variants of the basic preferential attachment
mechanism have also been proposed, and some of these lead
to changes in the values
of the exponents in the resulting power laws.
For extensive reviews of
work in this area, see Albert and Barab\'{a}si \cite{BA-RMP02},
Dorogovtsev and Mendes \cite{DorogMendes02}, and Newman
\cite{MEJN-SIAM}; for a survey of the rather
limited amount of mathematical work see \cite{BolloRior02}.
Most of this work concerns network models without reference to
an underlying geometric space.  Nor do
most of these models allow for heterogeneity of nodes, or
address physical constraints
on the capacity of the nodes.  Thus, while such models may
be quite appropriate for geometry-free networks, such as the web graph,
they do not seem to be ideally suited to the description of other
observed networks,
{\em e.g.}, the Internet graph.

In this paper,
instead of assuming preferential attachment, we show that it can arise
from a more basic underlying process, namely competition between
opposing forces.
The idea that power laws can arise from competing effects, modeled as
the solution of optimization problems with complex objectives, was proposed
originally by Carlson and Doyle \cite{CD-HOT99}.  Their ``highly
optimized tolerance'' (HOT) framework has reliable design as a primary
objective.  Fabrikant, Koutsoupias and Papadimitriou (FKP) \cite{FKP02}
introduce an elegant network growth model with such a mechanism, which
they called ``heuristically optimized trade-offs''.  As in many growth
models, the FKP network is grown one node at a time, with each new node
choosing a previous node to which it connects.  However, in contrast to
the standard preferential attachment types of models, a key feature of the
FKP model is the underlying geometry.  The nodes are points chosen
uniformly at random from some region, for example a unit square in the plane.
The trade-off is between the geometric consideration that it is desirable
to connect to a nearby point, and a networking consideration, that it
is desirable to connect to a node that is ``central'' in the network
as a graph.  Centrality is measured by using, for example, the graph
distance to the initial node. The model has a tunable, but fixed,
parameter, which determines the relative weights given to
the geometric distance and the graph distance.

The suggestion that competition between two metrics could be an
alternative to preferential attachment for generating power law degree
distributions represents an important paradigm shift.  Though FKP
introduced this paradigm for network growth, and FKP networks have
many interesting properties,
the resulting distribution is not a power law in
the standard sense \cite{BBBCR03}.  Instead the overwhelming majority
of the nodes are leaves (degree one), and a second substantial
fraction, heavily connected ``stars'' (hubs), producing a node degree
distribution which has clear bimodal features.%
\footnote{In simulations of the FKP model, this can be clearly
discerned by examining the probability distribution function (pdf);
for the system sizes amenable to simulations, it is less prominent
in the cumulative distribution function (cdf).}

Here, instead of directly producing power laws as a consequence of
competition between metrics, we show that such competition can give
rise to the preferential attachment mechanism, which in turn gives
rise to power laws.  Moreover, the power laws we generate have an upper
cutoff, which is more realistic in the context of many applications.



%% file: intro2.tex
\subsection{Overview of Competition-Induced Preferential Attachment}

We begin by formulating a general competition model for network growth.
Let $x_0, x_1, \dots, x_t$ be a sequence of random variables with values
in some space $\Lambda$.  We think of the points $x_0, x_1, \dots, x_t$
arriving one at a time
according to some stochastic
process.  For example, we typically take $\Lambda$ to be a compact
subset of $\mathbb R^d$, $x_0$ to be a given point, say the origin,
and $x_1, \dots, x_t$ to be i.i.d. uniform on $\Lambda$.  The network
at time $t$ will be represented by a graph,  $G(t)$,
on $t+1$ vertices, labeled $0,1,\dots, t$,
and at each time step, the new node attaches to one or several
nodes in the existing network.
For simplicity, here we assume that each new node connects
to a single node, resulting in $G(t)$ being
a tree.

Given $G(t-1)$, the new node, labeled $t$,
attaches to that node $j$ in the existing
network that minimizes a certain cost function representing the
trade-off of two competing effects, namely
connection or startup cost, and routing or performance cost.
The connection cost is represented
by a metric,
$ g_{ij}(t)$,
on $\{0, \dots, t\}$ which depends on $x_0, \dots, x_{t}$, but not
on the current graph $G(t-1)$, while the routing cost is
represented by a function,
$h_j(t-1)$,
on the nodes which depends on the current graph, but not
on the physical locations  $x_0,\dots,x_t$ of the nodes
$0,\dots,t$.  This leads to the cost function
\begin{equation}
c_{t} = {\min_j}\left[ \alpha g_{tj}(t)  + h_j(t-1) \right],
\label{cost}
\end{equation}
where $\alpha$ is a constant which determines the relative weighting
between connection
and routing costs.  We think of the function $h_j(t-1)$ as measuring
the centrality of the node $j$; for simplicity, we take it to be
the hop distance along the graph $G(t-1)$ from $j$
to the root $0$.

When $\Lambda$ is equipped with an appropriate norm $\| \cdot\|$, we
can use a simplified algorithm,
minimizing the cost only over those points $j$ which are closer to the
root than is the new  point:
\begin{equation}
\tilde c_t = {\min_{j: \| x_j \| < \| x_{t} \|}}
\left[ \alpha g_{tj}(t)  + h_j(t-1) \right].
\label{greedy-cost}
\end{equation}
In the  original FKP model, $\Lambda$
is a compact subset of $\mathbb R^2$, say the unit square, and the
points $x_i$ are independently uniformly distributed  on $\Lambda$.
The cost
function is of the form \eqref{cost}, with $g_{ij} =
d_{ij}$, the Euclidean metric (modeling the cost of
building the physical transmission line),
and $h_j(t)$ is the hop distance along the existing
network $G(t)$ from $j$
to the root.  A rigorous analysis of the degree distribution
of this two-dimensional model was
given in \cite{BBBCR03}, and the analogous one-dimensional
problem was treated in \cite{KenyonSchab}.

Our model is defined as follows.

\begin{defn}[ Border Toll Optimization Process]
Let $x_0=0$, and let $x_1,x_2,\dots$ be i.i.d., uniformly at
random in the unit interval $\Lambda = [0,1]$, and let
$G(t)$ be the following process: At $t=0$, $G(t)$ consists
of a single vertex $0$, the root.  Let $h_j(t)$ be the hop
distance to $0$ along $G(t)$,
and let $g_{ij}(t) = n_{ij}(t)$ be the number of
existing nodes between $x_i$ and $x_j$ at time $t$,
which we refer to as the
{\em jump cost} of $i$ connecting to $j$.
Given $G(t-1)$ at time $t-1$, a new vertex, labeled
$t$, attaches to the node $j$ which minimizes the cost function
\eqref{greedy-cost}.
Furthermore, if there are several nodes
$j$ that minimize this cost function and satisfy the constraint, we
choose the one whose position $x_j$ is nearest to $x_t$.
The process so defined is called the {\em border toll optimization process (BTOP)}.
\end{defn}

As in the FKP model, the routing cost
is just the hop distance to the root along the existing network.
However, in our model the
connection cost metric  measures the number of
``borders'' between two nodes: hence the name BTOP.  Note the
correspondence to the Internet, where the principal connection
cost is related to the number of AS domains crossed --
representing, {\em e.g.}, the overhead associated with BGP,
monetary costs of peering agreements, etc.
In order to facilitate
a rigorous analysis of our model, we took the simpler cost
function \eqref{greedy-cost}, so that the new node always attaches to
a node to its left.

It is interesting to note that the ratio of
the BTOP connection cost metric to that of the one-dimensional FKP model is
just the local density of nodes:  $n_{ij}/d_{ij} = \rho_{ij}$.
Thus the transformation between the two models is
equivalent to replacing the constant parameter $\alpha$ in the
FKP model with a variable parameter $\alpha_{ij}
= \alpha\rho_{ij}$ which changes as the network evolves in
time.  That $\alpha_{ij}$ is proportional to the local density
of nodes in the network reflects a model with an increase in
cost for local resources that are scarce or in high demand.
Alternatively, it can be thought of as reflecting the
economic advantages
of being first to market.

Somewhat surprisingly, the BTOP is equivalent
to a special case of the following process, which closely parallels
the preferential attachment model and makes no reference to any
underlying geometry.

\begin{defn}[Generalized Preferential Attachment with Fertility and Aging]
\label{def:cipa}
Let $A_1,A_2$ be two positive integer-valued parameters.
Let $G(t)$ be the following Markov process, whose states
are finite rooted trees in which each node is labeled
either {\em fertile} or {\em infertile}.  At time
$t=0$, $G(t)$ consists of a single fertile vertex.
Given the graph at time $t$, the new graph is formed in
two steps: first, a new vertex, labeled $t+1$ and initialized as
infertile,
connects to an old vertex $j$ with probability
zero if $j$ is infertile, and with probability
\begin{equation}
\label{cipa-pr}
Pr(t+1 \rightarrow j) = \frac{\min\{d_j(t),A_2\}}{W(t)}
\end{equation}
if $j$ is fertile.  Here, $d_j(t)$ is equal to $1$ plus the
out-degree of $j$, and $W(t)=\sum'_j \min\{d_j(t),A_2\}$, where the
sum is only over fertile vertices. Second,
if after the first step, $j$ has more than $A_1-1$ infertile
children, one of them, chosen uniformly at random, becomes
fertile. The process so defined is called a {\em generalized
preferential attachment process with fertility threshold $A_1$ and
aging threshold $A_2$.}
The special case $A_1=A_2$ is called the {\em competition-induced
preferential attachment process} with parameter $A_1$.
\end{defn}

The last definition is motivated by the following theorem,
to be proved in Section~\ref{sec:themodel}.

\begin{thm}  \label{thm:equiv}
The border toll optimization process
is equivalent to a the competition-induced preferential attachment
process with parameter $A= \lceil \alpha^{-1} \rceil.$
\end{thm}

Certain other limiting cases of the generalized
preferential attachment process are worth noting.
If $A_1 = 1$ and $A_2 = \infty$, we recover the
standard Barab\'{a}si-Albert model of preferential attachment.  If
$A_1 = 1$ and $A_2$ is finite, the model is equivalent to the standard
model of preferential attachment with a cutoff.
On the other hand, if $A_1 = A_2 = 1$, we get
a uniform attachment model.

The degree distribution of our random trees is characterized by
the following theorem, which asserts that almost surely (a.s.) the
fraction of vertices having degree $k$ converges to a specified
limit $q_k$, and moreover that this limit obeys a power law for $k
< A_2$, and decays exponentially above $A_2$.

\begin{thm} \label{thm:main-thm}
Let $A_1$, $A_2$ be positive integers and let $\tilde G(t)$
be the generalized preferential attachment process with
fertility parameter $A_1$ and aging parameter $A_2$.
Let $N_0(t)$ be the number of infertile vertices at time $t$,
and let $N_k(t)$ be the number of fertile vertices with
$k-1$ children at time $t$, $k\geq 1$.
Then:
\begin{enumerate}
\item \label{mthm:convergence}
There are numbers $q_k\in [0,1]$ such that,
for all $k\geq 0$
\begin{equation}
\label{mthm:empconv}
\frac{N_k(t)}{t+1}\to q_k\quad\text{a.s., as }t\to\infty.
\end{equation}
\item \label{mthm:qk-formula}
There exists a number $w \in [0,2]$ such that the $q_k$
are determined by the following equations:
\begin{eqnarray}
\label{mthm:1}
q_i & = & \left(\prod_{k=2}^i\frac{k-1}{k+w}\right)q_1
\quad\text{if}\quad i\leq A_2, \\
\label{mthm:2}
q_i & = & \left(\frac{A_2}{A_2+w}\right)^{i-A_2}q_{A_2}
\quad\text{if}\quad i> A_2
\end{eqnarray}
\[
1 = \sum_{i=0}^\infty q_i ,
\qquad\text{and}\qquad
q_0  =  \sum_{i=1}^\infty q_i\min\{i-1,A_1-1\}.
\]
\item \label{mthm:powerlaw}
There are positive constants
$c_1$ and $C_1$, independent of $A_1$ and $A_2$,
such that
\begin{equation} \label{eqn:mainthm_powerlaw}
c_1 k^{-(w+1)} < q_k/q_1 < C_1 k^{-(w+1)}
\end{equation}
for $1 \le k \le A_2$.
\item \label{mthm:monotonicity}
If $A_1=A_2$, the parameter $w$ is equal to $1$, and for general
$A_1$ and $A_2$, it decreases with increasing $A_1$, and increases with
increasing $A_2$.
\end{enumerate}
\end{thm}

Equation (\ref{eqn:mainthm_powerlaw}) clearly defines
a power law degree distribution
with exponent $\gamma = w+1$ for $k \leq A_2$.
Note that for measurements of the Internet the value of the exponent
for the power law is $\gamma \approx 2$.  In our border toll
optimization model,
where $A_1 = A_2$, we recover $\gamma=2$.

The convergence claim of Theorem~\ref{thm:main-thm} is proved
using a novel method which we believe is one of the main technical
contributions of this paper.  For preferential attachment models
which have been analyzed in the
past~\cite{ACL02,BBCR03,BRST01,CF01}, the convergence was
established using the Azuma-Hoeffding martingale inequality.  To
establish the bounded-differences hypothesis required by that
inequality, those proofs employed a clever coupling of the random
decisions made by the various edges, such that the decisions made
by an edge $e$ only influence the decisions of subsequent edges
who choose to imitate $e$'s choices.  A consequence of this
coupling is that if $e$ made a different decision, it would alter
the degrees of only finitely many vertices.  This in turn allows
the required bounded-differences hypothesis to be established.  No
such approach is available for our models, because the coupling
fails.  The random decisions made by an edge $e$ may influence the
time at which some node $v$ crosses the fertility or
attractiveness threshold, which thereby exerts a subtle influence
on the decisions of {\em every} future edge, not only those who
choose to imitate $e$.

Instead we introduce a new approach based on the second moment method.
The argument establishing the requisite second-moment upper bound is
quite subtle; it depends on a computation involving the eigenvalues of
a matrix describing the evolution of the degree sequence in a
continuous-time version of the model.  The key observation is that, in
this continuous-time model, the expected number of vertices of each
degree grows exponentially at a rate determined by the largest
eigenvalue, $w$, of this matrix, while the variance of the number of
vertices of each degree has an exponential growth rate which is at
most the second eigenvalue.  For the matrix in question, the top
eigenvalue has multiplicity 1, thus ensuring that the variance grows
more slowly than the mean.  We then translate this continuous-time
result into a rigorous convergence result for the original
discrete-time system.

%% file: equivalence.tex
\section{Equivalence of the two models}
\label{sec:themodel}

\subsection{Basic properties of the border toll optimization
process}

In this section we will turn to the BTOP
defined in the introduction,
establishing some basic properties which will enable
us to prove that it is equivalent to the competition-induced
preferential attachment model.
In order to avoid complications we exclude the case that some of
the $x_i$'s
are identical, an event that has probability zero.  We say that
$j\in\{0,1\dots,t\}$ lies to the right of $i\in\{0,1\dots,t\}$ if
$x_i<x_j$, and  we say that $j$ lies directly to the right of $i$
if $x_i<x_j$ but there is no $k\in\{1,\dots,t\}$ such that
$x_i<x_k<x_j$.  In a similar way, we say that $j$ is the first
vertex with a certain property to the right of $i$ if $j$ has that
property and there exists no $k\in\{1,\dots,t\}$ such that
$x_i<x_k<x_j$ and $k$ has the property in question.

\begin{defn}
A vertex $i$ is called {\em fertile at time $t$}
if a new point that arrives at time $t+1$
and lands directly to the right of $x_i$ attaches itself
to the node $i$.  Otherwise $i$ is called {\em infertile} at time
$t$.
\end{defn}

\begin{figure}[bct]
\vskip-1cm
{\hfill
\resizebox{4.5in}{!}{\includegraphics{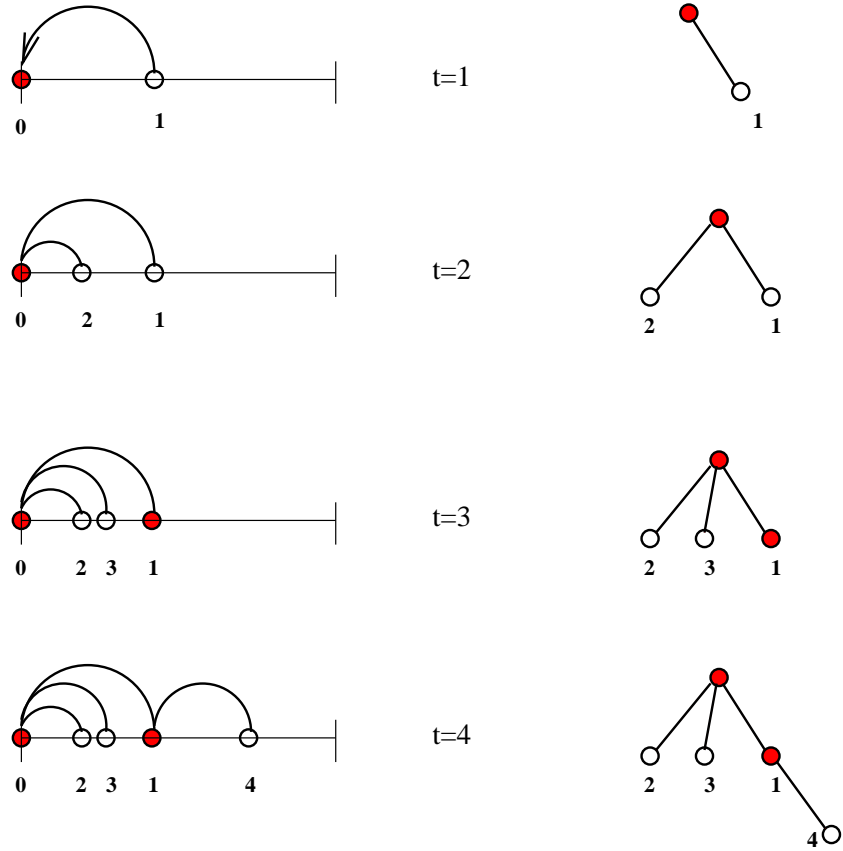}}\hfill}
\vspace{-0.2in}
\caption{A sample instance of BTOP for $A=3$, showing the process on the
unit interval (on the left), and the resulting tree (on the right).
Fertile vertices are marked red, infertile ones are marked white.
Note that vertex $1$ became fertile at $t=3$.}
\label{fig:tree-tA+1}
\end{figure}

This definition  is illustrated in Fig.~\ref{fig:tree-tA+1}.

\begin{lemma}
\label{lem:fertility}
Let $0<\alpha<\infty$, let $A=\lceil \alpha^{-1}\rceil$,
and let $0<t<\infty$.  Then

i) The node $0$ is fertile at time $t$.

ii) Let $i$ be fertile at time $t$.  If $i$
is the right most fertile vertex at time $t$
(case 1),
let $\ell$ be the number of infertile vertices
to the right of $i$.  Otherwise (case 2), let $j$ be the
next fertile vertex to the right of $i$, and let
$\ell=n_{ij}(t)$.  Then $0\leq \ell\leq A-1$,
and the $\ell$ infertile vertices located directly to the
right of $i$ are children of $i$. In case 2, if $h_{j}>h_i$, then
$j$ is a fertile child of $i$ and $\ell=A-1$.
As a consequence, the hop count between two consecutive
fertile vertices never increases by more than $1$ as we
move to the right, and if it increases by $1$, there
are $A-1$ infertile vertices between the two fertile
ones.

iii) Assume that the new vertex at time $t+1$ lands
between two consecutive fertile vertices $i$ and $j$,
and let $\ell=n_{ij}(t)$.  Then $t+1$ becomes a child of $i$.  If
$\ell+1<A$, the new vertex is infertile at time $t+1$, and the fertility
of all old vertices is unchanged.  If $\ell+1=A$
and the new vertex lies directly to the left of $j$,
the new vertex is fertile at time $t+1$ and the fertility
of the old vertices is unchanged.  If $\ell+1=A$
and the new vertex lies not directly to the left of $j$,
the new vertex is infertile at time $t+1$, the vertex
directly to the left of $j$ becomes fertile, and the fertility
of all other vertices is unchanged.

iv) If $t+1$ lands to the right of the
right most fertile vertex at time $t$, the statements
in iii) hold with $j$ replaced by the right endpoint of the interval
$[0,1]$, and $n_{ij}(t)$ replaced by the number of vertices to the right
of $i$.

v) If $i$ is fertile at time $t$, it is still fertile at time $t+1$.

vi) If $i$ has $k$ children at time $t$, the $\ell=\min\{A-1,k\}$
left most of them are infertile at time $t$, and any others are fertile.
\end{lemma}

\begin{proof}
Statement i) is trivial, and statements v) and vi) follow immediately from
iii) and iv), so we are left with ii) --- iv).
We proceed by induction on $t$.  If ii) holds at time $t$,
and iii)and iv) hold for a new vertex arriving at time $t+1$,
ii) clearly also holds at time $t+1$.  We therefore only have to prove
that ii) at time $t$ implies iii) and iv) for a new vertex arriving at time
$t+1$.
 Using, in particular, the last statement of ii) as a key
ingredient, the proof is straightforward but lengthy.  It is not
worth reproducing here.  The interested reader can find it in
Appendix~\ref{app:fertility}.
\end{proof}

\subsection{Proof of Theorem~\ref{thm:equiv}}

In the BTOP, note that our cost function
\begin{equation} \label{eqn:costfunc-repeat}
{\rm min_j}\left[ \alpha n_{tj}(t) + h_j(t-1) \right],
\end{equation}
and hence the graph $G(t)$,
only depends on the order of the vertices $x_0,\dots,x_t$, and not on their
actual positions in the interval $[0,1]$.  Let $\vec\pi(t)$
be the permutation of $\{0,1,\dots,t\}$ which orders the vertices
$x_0,\dots,x_t$ from left to right, so that
\begin{equation}
x_0=x_{\pi_0(t)}<x_{\pi_1(t)}<\dots<x_{\pi_t(t)}.
\end{equation}
(Recall that the vertices $x_0,x_1,\dots,x_t$ are pairwise distinct with
probability one.)
We can consider a change of variables, from the $x$'s to the
length of the intervals between successive ordered vertices:
\begin{equation}
s_i(t) \equiv x_{\pi_{i+1}(t)} - x_{\pi_{i}(t)}
\quad\text{if}\quad 0 \le i \le t-1
\quad\text{and}\quad s_t(t)=1-x_{\pi_t(t)}.
\end{equation}
The lengths then obey the constraint: $\sum_{i=0}^t s_i =
1.$ The set of interval lengths, $\vec{s}(t)$ together with the set of
permutation labels $\vec{\pi}(t)=(\pi_0(t),\pi_1(t),\dots,\pi_t(t))$ is
an equivalent representation to the original set of position
variables, $\vec{x}(t)$.

Let us consider the process $\{\vec \pi(t)\}_{t\geq 1}$.  It is not hard to
show that this process is a Markov process, with the initial permutation
being the trivial permutation given by $\pi_i(1)=i$, and the permutation
at time $t+1$ obtained from $\vec\pi(t)$ by inserting the new point
$t+1$ into a uniformly random position.  More explicitly, the new
permutation $\vec\pi(t+1)$ is
obtained from $\vec\pi(t)$ by choosing $i_o\in\{1,\dots,t+1\}$
uniformly at random, and setting
\begin{equation}
\pi_i(t+1)=
\begin{cases}
\pi_i(t)      &\text{if}\quad i\leq i_0\\
t +1          &\text{if}\quad i=i_0\\
\pi_{i-1}(t)&\text{if}\quad i>i_0.
\end{cases}
\end{equation}
Indeed, let $I_k(t)=[x_{\pi_k(t)},x_{\pi_{k+1}(t)}]$, and consider for a moment
the process $(\vec\pi(t),\vec s(t))$.  Then the
conditional probability that the next point arrives in the $k$-th
interval, $I_k$, depends only on the interval length at time $t$:
\begin{eqnarray}
Pr \left[ x_{t+1} \in I_k \left| \vec\pi(t), \vec{s}(t), \vec\pi(t-1),
\vec{s}(t-1), \dots, \vec\pi(0), \vec{s}(0) \right.\right] \nonumber \\
 = Pr \left[x_{t+1} \in I_k \left| \vec\pi(t), \vec{s}(t) \right.\right] = s_k(t).
\end{eqnarray}
Integrating out the dependence on the interval length from the above
equation we get:
\begin{eqnarray}
Pr \left[ x_{t+1} \in I_k \left| \vec\pi(t) \right.\right]
& = & \int Pr \left[ x_{t+1} \in I_k \left| \vec\pi(t), \vec{s}(t) \right.\right]
dP(\vec{s}(t)) \nonumber \\
& = & \int s_k(t) dP(\vec{s}(t)) = \frac{1}{t+1},
\end{eqnarray}
since after the arrival of $t$ points, there exist $(t+1)$ intervals.
The probability that the next point arrive in the $k$-th
interval is uniform over all the intervals, proving that
$\vec\pi(t)$ is indeed a Markov chain with the transition probabilities
described above.

With the help of Lemma~\ref{lem:fertility}, we now easily derive a description
of the graph $G(t)$ which does not involve any optimization problem.
To this end, let us consider a vertex $i$ with $\ell$ infertile children
at time $t$.  If a new vertex falls into the interval directly to the right
of $i$, or into one of the intervals directly to the right of an infertile
child of $i$, it will connect to the vertex $i$.  Since there is a total
of $t+1$ intervals at time $t$, the probability that a vertex $i$ with
$\ell$ infertile children grows an offspring is $(\ell+1)/(t+1)$.
By Lemma~\ref{lem:fertility} (vi), this number is equal to
$\min\{A,k_i\}/(t+1)$, where $k_i-1$ is the number of children of $i$.
Note that fertile children don't contribute to this
probability, since vertices falling into an interval directly to the right
of a fertile child will connect to the child, not the parent.

Assume now that $i$ did get a new offspring, and that it had
$A-1$ infertile children at time $t$.  Then the new vertex
is  either born fertile, or makes one of its
infertile siblings
fertile.
 Using the principle of deferred decisions,
 we may assume that with probability $1/A$ the new vertex becomes
 fertile,
and with probability $(A-1)/A$ an old one, chosen uniformly at random among the
$A-1$ candidates, becomes fertile.

We thus have shown that the solution $G(t)$ of
the optimization problem \eqref{eqn:costfunc-repeat}
can alternatively be described by the
competition-induced preferential attachment model
with parameter $A$.

%% file: convexp.tex
\section{Convergence of the Degree Distribution}
\label{sec:a1eqa2}

\subsection{Overview}

To characterize the behavior of the degree distribution, we will
derive a recursion which governs the evolution of the expected number
of vertices of each degree, at the time when there are $\tau$ nodes in
the network.  The coefficients of this recursion are random
variables depending on $W(\tau)$, the combined attractiveness
of all vertices at time $\tau$.  To simplify the analysis of
the recursion, we introduce a continuous-time model which is
equivalent to the original discrete-time model up to a (random)
reparametrization of the time coordinate.  The kernel of this
continuous-time Markov chain is a matrix $M$ whose coefficients
we identify explicitly.  In this section we will prove that
the {\em expected} degree distribution converges to a scalar multiple
of the eigenvector
$\hat{p}$ of $M$ associated with the largest eigenvalue $w$.  The
much more difficult proof that the {\em empirical} degree distribution
converges a.s. to the same limit is deferred to Appendix~\ref{appdx:convergence}.

\subsection{Notation}\label{subseq:notation}

Let $A\geq\max(A_1,A_2)$. At (discrete) time $\tau$, let
$N_0(\tau)$ be the number of infertile vertices at time $\tau$,
and,  for $k\geq 1$, let  $N_k(\tau)$ be the number of fertile
vertices with $k-1$ children at time $\tau$.  Let
$\tilde N_A(\tau)=N_{\geq A}(\tau)=\sum_{k\geq A}N_k(\tau)$,
and $\tilde N_k(\tau)=N_k(\tau)$ if $k<A$.
Finally let
$\hN_k(\tau) = \frac{1}{\tau+1} \tilde N_k(\tau)$,
and let $n_k(\tau), \hn_k(\tau)$
to be the expected values of $N_k(\tau), \hN_k(\tau)$,
respectively.

\subsection{Evolution of the expected value}
\label{subseq:evolexp}

From the definition of the generalized
preferential attachment model, it is easy to derive
the probabilities for the various alternatives which may
happen upon the arrival of the $(\tau+1)$-st node:
\begin{itemize}
\item  With probability $A_2 \tilde N_A(\tau)/W(\tau)$, it attaches to a node
of degree $\geq A$.  This increments $\tilde N_1$, and leaves
$\tilde N_A$ and all $\tilde N_j$ with $1<j<A$ unchanged.
\item  With probability $\min(A_2,k) \tilde N_k(\tau)/W(\tau)$, it attaches to
a node of degree $k$, where $1 \leq k < A$.
This increments $\tilde N_{k+1}$,
decrements $\tilde N_k$, increments $\tilde N_0$ or $\tilde N_1$
depending on whether
$k < A_1$ or $k \geq A_1$, and leaves all other $\tilde N_j$
with $j<A$ unchanged.
\end{itemize}
It follows that the discrete-time process
$\{\tilde N_k(\tau)\}_{k=0}^{\infty}$
is equivalent to the state of the following continuous-time stochastic
process (with time parameter $t$) at the time of the $\tau$-th event.
\begin{itemize}
\item   With rate $\tilde A_2 N_A(t)$, $\tilde N_1$ increases by $1$.
\item   For every $0<k<A$, with rate
$\tilde N_k(t)\min(k,A_2)$, the following happens:
\[
\tilde N_k\to \tilde N_k-1\ \ \  ; \ \ \ \tilde N_{k+1}\to \tilde N_{k+1}+1\ \ \   ;
\ \ \ \tilde N_{g(k)} \to \tilde N_{g(k)}+1
\]
where $g(k)=0$ for $k<A_1$ and $g(k)=1$ otherwise.
\end{itemize}
Let $M$ be the following $A\times A$ matrix:
\[
M_{i,j}=\left\{
\begin{array}{ll}
-\min(j,A_2) & \mbox{if } \ 1\leq i=j\leq A-1 \\
\min(j,A_2) & \mbox{if } \ 1\leq i=j+1\leq A \\
\min(j,A_2) & \mbox{if } \ j=1 \mbox{ and } i>A_1 \\
0 & \mbox{otherwise}.
\end{array}
\right.
\]
Then, for the continuous time process, for every $t>s$,
the  conditional expectations of the vector
$\tilde N(t)$ are given by
\begin{equation}\label{eq:exval}
E\left(\tilde N(t)|\tilde N(s)\right)=e^{(t-s)M}\tilde N(s).
\end{equation}
It is easy to see that the matrix $e^M$ has all positive entries, and
therefore (by the Perron-Frobenius Theorem) $M$
has a unique eigenvector $\hat{p}$ of $\ell_1$-norm 1,
having all positive entries.
Let $w$ be the eigenvalue corresponding to $\hat{p}$. Then $w$ is real,
it has multiplicity $1$, and it exceeds the real part of
every other eigenvalue.
Therefore, for every non-zero vector $n$ with non-negative entries,
\[
\lim_{t\to\infty}e^{-tw}e^{tM}n=\langle\hat{a},n\rangle\hat{p}
\]
where $\hat{a}$ is the eigenvector of $M^{\trans}$ corresponding
to $w$. Note that $\langle\hat{a},n\rangle>0$ because $n$ is
non-zero and non-negative, and $\hat{a}$ is positive, again by
Perron-Frobenius. Therefore, up to a scalar factor, the vector
$\hn(t) := E\left(e^{-tw} \tilde N(t)\right)$ converges to
$\hat{p}$ as $t \rightarrow \infty$.  Note that this implies, in
particular, that $w>0$.  We can also show 
that $w\leq 2$.  This is a consequence of
Claim~\ref{claim:couple} in Appendix~\ref{appdx:convergence},
which says that $\hN_k(t)$ is stochastically dominated by
a stochastic process $X_t$ satisfying $E(X_t) \sim e^{2t}$.

 To conclude that the discrete time
version,  $\hn(\tau)$, converges to $\hat p$ as well, one needs
show that, a.s., $\tau$ is finite for all finite $t$.  This is
done in Claims~\ref{claim:spenc} and~\ref{claim:couple} in
Appendix~\ref{appdx:convergence}.

%% file: powerlaw.tex
\section{Power law with a cutoff}
\label{sec:powerlaw}

In the previous section, we saw that for every $A>\max\{A_1,A_2\}$, the limiting proportions up to $A-1$ are $\hat{p}$ where $\hat{p}$
is the eigenvector corresponding to the highest eigenvalue $w$ of the
$A$-by-$A$ matrix
\begin{equation} \label{eqn:M}
M_{i,j}=\left\{
\begin{array}{ll}
-\min(j,A_2) & \mbox{if } \ 1\leq i=j\leq A-1 \\
\min(j,A_2) & \mbox{if } \ 1\leq i=j+1\leq A \\
\min(j,A_2) & \mbox{if } \ i=1 \mbox{ and } j>A_1 \\
0 & \mbox{otherwise}.
\end{array}
\right.
\end{equation}
Therefore, the proportions $p$ satisfy the equation:
\begin{eqnarray}\label{eq:rec}
wp_i=-\min(i,A_2)p_i + \min(i-1,A_2)p_{i-1} \ \ \ \ \ \ i \geq 2
\end{eqnarray}
where the normalization is determined by $\sum_{i=1}^Ap_i=1$.
From (\ref{eq:rec}) we get that for $i\leq A_2$,
\begin{equation}\label{eq:powreg}
p_i=\left(\prod_{k=2}^i\frac{k-1}{k+w}\right)p_1
\end{equation}
and for $i>A_2$
\begin{equation}\label{eq:exreg}
p_i=\left(\frac{A_2}{A_2+w}\right)^{i-A_2}p_{A_2}
\end{equation}
Clearly, (\ref{eq:exreg}) is exponentially decaying.
There are many ways to see that (\ref{eq:powreg}) behaves like a power-law
with degree $1+w$.  The simplest would probably be:
\begin{eqnarray}\label{eq:jifa}
\frac{p_i}{p_1}=\left(\prod_{k=2}^i\frac{k-1}{k+w}\right)
&=&\exp\left(\sum_{k=2}^i\log\left(\frac{k-1}{k+w}\right)\right)\\
\nonumber
=\exp\left(\sum_{k=2}^i\left(\frac{-1-w}{k+w}\right)+O(1)\right)
&=&\exp\left((-1-w)\left(\sum_{k=2}^i(k+w)^{-1}\right)+O(1)\right)\\
\nonumber
=\exp\left((-1-w)\left(\sum_{k=2}^ik^{-1}\right)+O(1)\right)
&=&\exp\left((-1-w)\left(\sum_{k=2}^i\log\left(\frac{k+1}{k}\right)\right)+O(1)\right)\\
\nonumber
=\exp\left((-1-w)\log(i/2)+O(1)\right)&=&O(1)i^{-1-w}.
\end{eqnarray}
Note that the constants implicit in the $O(\cdot)$ symbols
do not depend on $A_1$, $A_2$ or $i$,
due the fact that $0<w\leq 2$.
(\ref{eq:jifa}) can be stated in the following way:

\begin{prop} \label{prop:power-law}
There exist $0<c<C<\infty$ such that for every $A_1$, $A_2$ and $i\leq A_2$,
if $w=w(A_1,A_2)$ is as in (\ref{eq:rec}), then
\begin{equation} \label{eq:PowerLawForP}
ci^{-1-w}\leq\frac{p_i}{p_1}\leq Ci^{-1-w}.
\end{equation}
\end{prop}
The vector $q = (q_1,q_2,\ldots)$ is a scalar multiple of $\hat{p}$,
so equations (\ref{mthm:1}), (\ref{mthm:2}), and (\ref{eqn:mainthm_powerlaw})
in Theorem~\ref{thm:main-thm} (and the comment immediately following it)
are consequences of equations
(\ref{eq:powreg}), (\ref{eq:exreg}), and (\ref{eq:PowerLawForP})
derived above.  It remains to prove the normalization conditions
\[
\sum_{i=0}^{\infty} q_i  =  1; \qquad
q_0  =  \sum_{i=1}^{\infty} q_i \min(i-1, A_1-1)
\]
stated in Theorem~\ref{thm:main-thm}.  These follow from the equations
\[
\sum_{i=0}^{\infty} N_i(t)  =  t+1; \qquad
N_0(t)  =  \sum_{i=1}^{\infty} N_i(t) \min(i-1, A_1-1).
\]
The first of these simply says that there are $t+1$ vertices at
time $t$; the second equation is proved by counting the number
of infertile children of each fertile node.

The monotonicity properties of $w$ asserted in
part~\ref{mthm:monotonicity} of Theorem~\ref{thm:main-thm}
are proved in Appendix~\ref{appdx:monotonicity}.
The concentration of the empirical degree distribution
is proved in Appendix~\ref{appdx:convergence}.

%% file: appdx1.tex
\section{Proof of Lemma~\ref{lem:fertility}}
\label{app:fertility}

In this appendix, we complete the proof of Lemma~\ref{lem:fertility}.

To this end, let us first recall that the only non-trivial part is
the fact that ii) at time $t$ implies iii) and iv) for a new vertex arriving at time
$t+1$.
Assume thus that ii) holds at time $t$.

At time $t+1$, a new vertex arrives,
and falls directly to the right of some vertex $k$.
Let $i$ be the nearest vertex to the left of $k$ that was
fertile at time $t$
(if $k$ is fertile at time $t$, we set $i=k$) and let $j$ be the
nearest vertex to the right of $i$ that was fertile at time
$t$ (we assume for the moment that $i$ is not the right most
fertile vertex at time $t$), let $\ell$ be the number
of vertices between $i$ and $j$ at time $t$.

Let us first prove that the vertex $t+1$ connects to $i$.
If $i=k$, this is obvious, since $i$ is fertile at time
$t$. We may therefore assume that
$k\neq i$.  For the new vertex $t+1$, the cost of connecting
to the vertex $i$ is then equal to
$\alpha (n_{ik}(t)+1)$.
Let us first compare this cost to the cost of connecting
to a fertile vertex $i'$ to the left of $i$.
Let $i_0=i'$, let $i_s=i$,
and let $i_1,\dots,i_{s-1}$ be the fertile vertices between
$i'$ and $i$, ordered from left to right.  If
$h_{i_{m-1}}<h_{i_m}$,
we use the inductive assumption ii) to conclude that
 the number of infertile vertices between $i_{m-1}$ and $i_m$ is
equal to $A-1$, and $h_{i_{m-1}}=h_{i_m}-1$.  A decrease of $q$ in the
hop cost is therefore accompanied by an increase in the
jump cost of at least $\alpha A q\geq q$.  As a consequence,
it never pays
to connect to a fertile vertex $i'$ to the left of $i$.  The cost
of connecting to an infertile vertex to the left of $i$ is even
higher, since the hop count of an infertile vertex is at best equal
to the hop count of the next fertile vertex to the right.  We
therefore only have to consider the connection cost to some of the
infertile children of $i$. But again, the hop count is worse by $1$
when compared to the hop count of $i$, and the jump cost is at
best reduced by $(A-1)\alpha<1$, proving that the cost of
connecting to $i$ is minimal.

To discuss the fertility of the vertices in the graph
$G(t+1)$, we need to consider the arrival of a second
vertex, labeled $t+2$.  If $t+2$ falls to the left of $t+1$,
it will face an optimization problem that has not been changed
by the arrival of the vertex $t+1$, implying that the fertility
of the vertices to the left of $t+1$ is unchanged.  If $t+2$ falls
to the right of $j$, the cost of connecting to $j$ or one of
the vertices to the right of $j$ is the same as before, and the
cost of connecting to a vertex to the left of $j$ is at best equal
(the cost of connecting to any vertex to the left of $t+1$
is in fact higher, due to the additional cost of jumping over
the vertex $t+1$).  Therefore, the vertex $t+2$ will still prefer
to connect to either $j$ or one of the vertices to the right of
$j$, implying that the fertility of the vertices to the right
of $j$ has not changed at all.  We therefore are left with
analyzing the case where $t+2$ falls between $t+1$ and $j$.
Again, the vertex $t+2$ will prefer $i$ over any vertex to the
left of $i$ (the cost analysis is the same as the one used
for $t+1$ above), so we just have to compare the costs
of connecting to the different vertices between $i$ and $j$.
If $\ell+1<A$, this will again imply that $t+2$ connect to
$i$; but if $\ell+1=A$, the vertex $t+2$ will only connect
to $i$ if it does not fall to the right of the right most
of the now $\ell+1$ vertices between $i$ and $j$.  If it
falls to the right of this vertex, it will be as expensive to
connect to the right most
of the now $\ell+1$ vertices between $i$ and $j$
as it is to connect to $i$.  Recalling out convention of
connecting to the nearest vertex to the left if there
is a tie in costs, this proves that now $t+2$ connects to
the right most vertex between $i$ and $j$, implying that
this vertex is fertile.

The above considerations prove the fertility statements in
iii), and thus completes the proof of iii).
The case where $i$ is the right most fertile vertex at time
$t$ is similar (in fact, it is slightly easier since it involves
less cases), and leads to the proof of iv).  This completes the
proof of Lemma~\ref{lem:fertility}

%% file: appdx2.tex
\section{Concentration of $\hN_k(t)$}
\label{appdx:convergence}

\subsection{Concentration of the continuous time process}
In order to show concentration of the continuous time process, we will prove the following two lemmas:
\begin{lemma}\label{lem:egrow}
For every $u<w$ and every $1\leq k\leq A$, a.s. for every $t$ large enough,
\[
\tilde N_k(t)>e^{ut}.
\]
\end{lemma}
and
\begin{lemma}\label{lem:difnotgrow}
There exists $v<w$ s.t. for every $1\leq k<j\leq A$ a.s. for every $t$ large enough,
\[
p_j\tilde N_k(t)-p_k\tilde N_j(t)<e^{vt}.
\]
\end{lemma}
In order to prove Lemmas \ref{lem:egrow} and \ref{lem:difnotgrow}, we need to use some estimates
considering the standard birth process described below.
\begin{defn}
Let $\{o_n\}_{n=1}^\infty$ be independent exponential random variables,
so that $\E(o_n)=\frac{1}{2}n^{-1}$.
For $t\in[0,\infty)$, let $X_t=\inf\{n:\sum_{k=1}^no_k>t\}$.
Then $X$ is called the {\bf standard birth process}.
\end{defn}
The following claim will be proved in Appendix \ref{app:prclaim}.
\begin{clm}\label{claim:spenc}
$X_t$ is almost surely finite for every $t$. Furthermore, there exists
a constant $C_s$ such that for every $t_2>t_1$ and $x$ and $k$,
\begin{equation}\label{eq:spenc}
\prob\left(\left.X_{t_2}>kxe^{2(t_2-t_1)}\right|X_{t_1}=x\right)<\frac{C_s}{x(k-1)^2}.
\end{equation}
\end{clm}
The standard birth process is connected to our discussion
through the following easy claim:
\begin{clm}\label{claim:couple}
Let $\|\tilde N(t)\|=\sum_{k=1}^A\tilde N_k(t)$. Let $T\geq 0$, let $x\geq y$,
and let $X$ be a standard birth process.
Then $\left\{\left.\{X_t\}_{t\geq T}\right|X_T=x\right\}$
stochastically dominates
$\left\{\left.\{\|\tilde N(t)\|\}_{t\geq T}\right|\|\tilde N(T)\|=y\right\}$.
\end{clm}
\begin{proof}
Let $\{r_n\}_{n=1}^{\infty}$ be i.i.d. exponential
random variables with mean 1.  Then $\sum_{k=1}^n o_k$
has the same distribution as $\sum_{k=1}^n r_k/2k$.
The time at which the $n$-th node is born
has the same distribution as $\sum_{k=1}^n r_k/W(k)$,
where $W(k)$ denotes the combined attractiveness of
all nodes at time $k$.  The claim follows now from
the observation that $W(k) \leq 2k$.
\end{proof}

\begin{proof}[Proof of Lemma \ref{lem:difnotgrow}]
We use a martingale to bound the variance. Fix $T$, and let
\[L_t=\E\left(\left.p_j\tilde N_k(T)-p_k\tilde N_j(T)\right|\tilde N(t)\right)\]
Clearly, $L_t$ is a (continuous time) martingale. Let $b=b^{(j,k)}$ be the vector
\[
b_i=\left\{
\begin{array}{ll}
-p_k & \mbox{if } i=j\\
p_j & \mbox{if } i=k\\
0 & \mbox{otherwise}.
\end{array}
\right.
\]
By (\ref{eq:exval}), we know that $L_t = b^{\trans} e^{M(T-t)}n(t).$
$J\hat{p}=0$, and therefore the norm of $Je^{M(T-t)}$ is bounded by $e^{(T-t)v^\prime}$ for some $v^\prime<w$. Without loss
of generality, we may assume that $v^\prime>w/2$.
\begin{clm}\label{claim:mati}
\begin{equation}\label{eq:secmom}
\var\left(p_j\tilde N_k(T)-p_k\tilde N_j(T)\right)<C\exp(2v^\prime T)
\end{equation}
For some constant $C$.
\end{clm}
\begin{proof}
Let $0<\epsilon<\exp(-10T)$ be such that $K=T/\epsilon$ is an integer number.
Then, $\{U_k=L_{k\epsilon}\}_{k=0}^{K}$ is a martingale, and
\[
\var\left(p_j\tilde N_k(T)-p_k\tilde N_j(T)\right)=\sum_{k=0}^{K-1}\var(U_{k+1}-U_k)
\]
We want to estimate the variance of $(U_{k+1}-U_k)$. Let $v_k=\|\tilde N_{(k+1)\epsilon}-\tilde N_{k\epsilon}\|$. Clearly,
\[
\var(U_{k+1}-U_k)\leq\var(v_k)\exp\left[2v^\prime(T-(k+1)\epsilon)\right]
\]
%
%
Using Claims \ref{claim:spenc} and \ref{claim:couple},
\begin{eqnarray*}
\var(v_k)&=&\E\left(\var\left(v_k\left|\tilde N_{k\epsilon}\right.\right)\right)+
\var\left(\E\left(v_k\left|\tilde N_{k\epsilon}\right.\right)\right)\\
&\leq& \exp(wk\epsilon)\left(e^{4\epsilon}-1\right)+
\exp(4k\epsilon)\left(e^{2\epsilon}-1\right)^2\\
&\leq& 5\epsilon\exp(wk\epsilon)+4\epsilon^2\exp(4k\epsilon)<C_0\epsilon\exp(wk\epsilon)
\end{eqnarray*}
for $C_0=6$, by the choice of $\epsilon$.
Therefore,
\begin{eqnarray*}
\var\left(p_j\tilde N_k(T)-p_k\tilde N_j(T)\right) &<&
C_0\epsilon\sum_{k=0}^{K-1}\exp\left(
wk\epsilon+2v^\prime(T-(k+1)\epsilon)
\right)\\
&\leq& C_0e^{2v^\prime T}\int_0^T e^{(w-2v^\prime)t}dt
<C\exp(2v^\prime T)
\end{eqnarray*}
for
\[
C=C_0\int_0^\infty e^{(w-2v^\prime)t}dt<\infty.
\]
\end{proof}
Choose some $v$ strictly between $v^\prime$ and $w$  in a way that $w-v<0.25\min(0.1,v-v^\prime)$
and let $\delta=\min(0.1,v-v^\prime)$. Using Chebyshev's inequality,
\begin{equation}\label{eq:frcheb}
\prob\left(p_j\tilde N_k(T)-p_k\tilde N_j(T)>\frac{1}{3}e^{vT}\right)\leq Ce^{-2\delta T}
\end{equation}
Let $\{T_i\}_{i=1,2,\ldots}$ be such that $e^{2\delta T_i}=i^2$.
By Borel-Cantelli, almost surely there exists $i_0$ such that for all $i>i_0$,
\begin{equation}\label{eq:somt}
p_j\tilde N_k(T_i)-p_k\tilde N_j(T_i)<\frac{1}{2}e^{vT_i}
\end{equation}
We want to show that almost surely for all $T$ large enough,
\begin{equation}\label{eq:alw}
p_j\tilde N_k(T)-p_k\tilde N_j(T)<e^{vT}
\end{equation}
We know that $\E(\tilde N(T_i))=O(\exp(wT_i))$, and
using a martingale argument similar to the one in Claim \ref{claim:mati}, we get that
$\var(\tilde N(T_i))=O(\exp(2wT_i))$ and therefore
\[
\prob(\tilde N(T_i)>e^{(w+0.6\delta)T_i})<C_le^{-1.2\delta T_i}=C_li^{-1.2}
\]
for some constant $C_l$,
and therefore, if $m(i)$ is the number of moves between $T_i$ and $T_{i+1}$, then
\begin{eqnarray}\label{eq:befbade}
\nonumber
&&\prob\left(m(i)>\frac{1}{2}e^{vT_i}\right)\\
\nonumber
&\leq&\prob\left(\tilde N(T_i)>e^{w+(0.6\delta)T_i}\right)
+\prob\left(\left.m(i)>\frac{1}{2}e^{vT_i}\right|\tilde N(T_i)\leq e^{w+(0.6\delta)T_i}\right)\\
&\leq& C_li^{-1.2}+C_se^{-(w+0.6\delta)T_i}
\end{eqnarray}
where the last inequality uses Claim \ref{claim:spenc} and the fact that
\[
\frac{1}{2}e^{vT_i}>2e^{(w+0.6\delta)T_i}(\exp(2(T_{i+1}-T_i))-1)
\]
Using
Borel-Cantelli, we conclude that almost surely,
\begin{equation}\label{eq:bade}
\sum_{k=1}^{A}\left|\tilde N_k(T)-\tilde N_k(T_i)\right|<\frac{1}{2}e^{{vT_i}}
\end{equation}
for all $k$ and all $i$ large enough and all $T$ between $T_i$ and $T_{i+1}$.
(\ref{eq:alw}) follows from (\ref{eq:bade}).
\end{proof}

\begin{proof}[Proof of Lemma \ref{lem:egrow}]
Using the same martingale argument as above, we can conclude that $\var(\tilde N_A(t)|\tilde N_A(0)=1)<C_1e^{2wt}$,
while $\E(\tilde N_A(t)|\tilde N_A(0)=1)>C_2e^{wt}$. Therefore there exists $\rho>0$ such that
\begin{equation}\label{eq:usecmom}
\prob\left(\tilde N_A(t)>\rho e^{wt}|\tilde N_A(0)=1\right)>\rho.
\end{equation}
Fix some large $T$, and let $t_i=iT$. Then using (\ref{eq:usecmom}) and independence,
\begin{equation}\label{eq:even1}
\prob\left(\left.\tilde N_A(t_i)>\frac{\rho^2}{2} e^{wT}\right|\tilde N_A(t_{i-1})\right)>1-e^{-\frac{1}{16}\tilde N_A(t_{i-1})}
\end{equation}
where (\ref{eq:even1}) was obtained using Chernoff's bound.
From (\ref{eq:even1}), we get that almost surely, for all $i$ large enough,
\[
\tilde N_A(t_i)>\exp\left(i\left[wT+\log\left(\frac{\rho^2}{2}
\right)\right]\right).
\]
$\tilde N_A(t)$ is monotone increasing, and therefore
\begin{equation}\label{eq:befuse}
\tilde N_A(t)>C\exp\left(t\left[w-\frac{1}{T}\log\left(\frac{\rho^2}{2}
\right)\right]\right).
\end{equation}
for all $t$ large enough. Using Lemma \ref{lem:difnotgrow} we conclude that
\[
\tilde N_k(t)>C\exp\left(t\left[w-\frac{1}{T}\log\left(\frac{\rho^2}{2}
\right)\right]\right)>e^{ut}
\]
for all $k$ and large enough $t$.
\end{proof}
\begin{prop}\label{prop:costcont}
For every $k$ and $j$, almost surely
\begin{equation}\label{eq:fuf}
\lim_{t\to\infty}\frac{\tilde N_k(t)}{\tilde N_j(t)}=\frac{p_k}{p_j}
\end{equation}
\end{prop}
\begin{proof}
This follows immediately from Lemma \ref{lem:egrow} and Lemma \ref{lem:difnotgrow}.
\end{proof}
\subsection{Back to discrete time}
\begin{prop}
For the discrete time process, and $A>\max\{A_1,A_2\}$
there exists a vector $\hat{q}$ such that, for $k\leq A$, we have
\begin{equation}
\label{conv-to-qk}
\lim_{\tau\to\infty}\frac{\tilde N_k(\tau)}{\tau+1}=q_k.
\end{equation}
\end{prop}
\begin{proof}
The $i$-th newcomer is of degree zero with probability
\[
\frac
{\sum_{k=1}^{A_1-1}\tilde N_k(\tau)}
{\sum_{k=1}^{A}\tilde N_k(\tau)}
\]
However, by (\ref{eq:fuf}),
this expression tends to a limit, and therefore, using the law of large numbers,
\begin{equation}\label{eq:valq0}
\lim_{\tau\to\infty}\frac{ N_0(\tau)}{\tau+1}=q_0=\frac
{\sum_{k=1}^{A_1-1}p_k}
{\sum_{k=1}^{A}p_k.}
\end{equation}
Using (\ref{eq:fuf}) once more, the proposition now follows for
$k\geq 1$ with $q_k=(1-q_0)p_k$.
\end{proof}

Note that the above proposition implies that $q_k$ and hence
$p_k$ is independent of $A$ if $A>k$, since the left hand side
of \eqref{conv-to-qk} does not depend on $A$ if $A>k$. So,
in particular, $p_1$ does not depend on $A$.

%% file: appdx3.tex
\section{Monotonicity properties of $w$}
\label{appdx:monotonicity}
In this section we will prove that the exponent $1+w$ of
the power law in Proposition~\ref{prop:power-law}
is monotonically decreasing in $A_1$ and monotonically
increasing in $A_2$.  For this purpose, it will be
useful to define a family of matrices, parametrized by
two vectors $\yvec, \zvec \in \R^n$, which
generalizes the matrix $M$ appearing in (\ref{eqn:M}),
whose top eigenvalue is $w$.

Given vectors $\yvec = (y_1,y_2,\ldots,y_n),
\zvec = (z_1,z_2,\ldots,z_n) \in \R^n$, let
$\Myz$ denote the $n$-by-$n$ matrix
whose $(ij)$-th entry is:
\[
M_{i,j}(\yvec,\zvec) = \left\{
\begin{array}{ll}
-y_j & \mbox{if } \ i=j  \\
y_j & \mbox{if } \ i=j+1 \\
z_j & \mbox{if } \ i=1 \mbox{ and } j>1 \\
0 & \mbox{otherwise}.
\end{array}
\right.
\]
Thus, for instance, the matrix $M$ defined in (\ref{eqn:M})
is $\Myz$, where
\begin{eqnarray*}
\yvec & = & (1,2,\ldots,A_2-1,A_2,A_2,A_2,\ldots,A_2) \\
\zvec & = & (0,0,\ldots,0,\min(A_1,A_2),
\min(A_1+1,A_2),\ldots,A_2,A_2)
\end{eqnarray*}
For the remainder of this section, we will assume:
\begin{eqnarray}
\label{eqn:yi}
& \bullet &  y_i > 0 \mbox{ for } 1 \leq i \leq n, \\
\label{eqn:zi}
& \bullet &  z_i \geq 0 \mbox{ for } 1 \leq i \leq n,  \\
\label{eqn:zn}
& \bullet & z_n > 0.
\end{eqnarray}
All of these criteria will be satisfied by the matrices
$\Myz$ which arise in proving the desired monotonicity
claim.  It follows from (\ref{eqn:yi}),(\ref{eqn:zi}),
and (\ref{eqn:zn}) that if we add a suitably large scalar
multiple of the identity matrix to $\Myz$, we obtain an
irreducible matrix $\Myz + B I$ with non-negative entries.
The Perron-Frobenius Theorem guarantees that $\Myz + BI$
has a positive real eigenvalue $R$ of multiplicity 1, such
that all other complex eigenvalues have modulus $\leq R$;
consequently $\Myz$ has a real eigenvalue $w = R-B$, of
multiplicity 1, such that the real part of every other
eigenvalue is strictly less than $w$.

We will study how $w$ varies under perturbations of the
parameters $\yvec,\zvec$.  Let $P(\lambda,\yvec,\zvec)$
be the characteristic polynomial of $\Myz$, i.e.
\[
P(\lambda,\yvec,\zvec) = \det(\lambda I - \Myz).
\]
This is a polynomial of degree $n$ in $\lambda$ (with
coefficients depending smoothly on $\yvec,\zvec$),
whose largest real root
$w(\yvec,\zvec)$ exists and has multiplicity 1, provided
$(\yvec, \zvec)$ belongs to the region $V \subset \R^n \times \R^n$
determined by (\ref{eqn:yi}),(\ref{eqn:zi}),
and (\ref{eqn:zn}).  It follows from the Implicit Function
Theorem that $w(\yvec,\zvec)$ is a smooth function of $(\yvec,\zvec)$
in $V$, satisfying:
\begin{equation} \label{eqn:IFT}
\left. \left( \frac{\partial P}{\partial y_i}
\, + \, \frac{\partial w}{\partial y_i} \cdot
\frac{\partial P}{\partial w} \right)
\right|_{(w,\yvec,\zvec)} = 0
\end{equation}
If $\xvec$ is any vector in $\R^n \times \R^n$, and
$\partial_{\xvec}$ is the corresponding directional
derivative operator, we have from (\ref{eqn:IFT}):
\begin{equation} \label{eqn:dirdiv}
\partial_{\xvec} w(\yvec,\zvec) =
- \frac{\partial_{\xvec} P(w,\yvec,\zvec)}
{(\partial P / \partial w)|_{(w,\yvec,\zvec)}}.
\end{equation}
We know that $(\partial P / \partial w)|_{(w,\yvec,\zvec)} > 0$
because $P$ is a polynomial with positive leading coefficient,
$w$ is its largest real root, and $w$ has multiplicity 1.  Thus
we've established:
\begin{clm} For any vector $\xvec \in \R^n \times \R^n$,
and any $(\yvec,\zvec) \in V$, put $w = w(\yvec,\zvec)$.  Then
the directional derivatives
$\partial_{\xvec} w(\yvec,\zvec)$ and
$\partial_{\xvec} P(w,\yvec,\zvec)$ have
opposite signs.
\label{claim:dirdiv}
\end{clm}
This allows monotonicity properties of $w$ to be deduced from
calculations involving directional derivatives of $P$.
Given the definition of $\Myz$, it is straightforward to
compute that
\begin{equation} \label{eqn:charpoly}
P(\lambda,\yvec,\zvec) \; = \; \det(\lambda I - \Myz) \; = \;
\prod_{i=1}^n (\lambda + y_i) \; - \;
\sum_{j=2}^n P_j(\lambda,\yvec,\zvec),
\end{equation}
where
\begin{equation} \label{eqn:pj}
P_j(\lambda,\yvec,\zvec) =
\left( \prod_{i=1}^{j-1} y_i \right)
z_j
\left( \prod_{i=j+1}^n (\lambda + y_j)
\right).
\end{equation}
The following three lemmas encapsulate the requisite
directional derivative estimates.
\begin{lemma} \label{lem:mono1}
$(\partial P / \partial z_k)|_{(w,\yvec,\zvec)} < 0$
for $(\yvec, \zvec) \in V$.
\end{lemma}
\begin{proof}
\[
\partial P / \partial z_k =
- \partial P_k / \partial z_k =
- \left( \prod_{i=1}^{k-1} y_i \right)
\left( \prod_{i=k+1}^n (w + y_i) \right) < 0.
\]
\end{proof}
\begin{cor}
\label{cor:wA1}
$w$ is monotonically decreasing in $A_1$.
\end{cor}
\begin{proof}
Increasing $A_1$ from $k$ to $k+1$ has no effect
on $\yvec$,
and its only effect on $\zvec$ is to decrease $z_k$
from $\min(k,A_2)$ to 0.  As we move in the $-z_k$
direction, the directional derivative of $P$ is
positive, so the directional derivative of $w$ is
negative by Claim~\ref{claim:dirdiv}.
Thus $w$ decreases as we increase $A_1$ from $k$
to $k+1$.
\end{proof}
\begin{lemma} \label{lem:mono2}
$(\partial P / \partial y_k)|_{(w,\yvec,\zvec)} < 0$
if $(\yvec,\zvec) \in V$ and $z_k=0$.
\end{lemma}
\begin{proof}
\begin{eqnarray*}
\frac{\partial P}{\partial y_k}
& = &
\frac{\partial}{\partial y_k} \left[
\prod_{i=1}^n (w+y_i) \right] \; - \;
\sum_{j=2}^n \frac{\partial P_j}{\partial y_k} \\
& = &
\frac{1}{w+y_k} \prod_{i=1}^n (w+y_i) \; - \;
\frac{1}{y_k} \sum_{j=2}^{k-1} P_j \; - \;
\frac{1}{w+y_k} \sum_{j=k+1}^n P_j \\
& < &
\frac{1}{w+y_k} \prod_{i=1}^n (w+y_i) \; - \;
\frac{1}{w+y_k} \sum_{j=2}^{k-1} P_j \; - \;
\frac{1}{w+y_k} \sum_{j=k+1}^n P_j \\
& = & \frac{P(w,\yvec,\zvec)}{w+y_k}  \\
& = & 0
\end{eqnarray*}
\end{proof}
\begin{lemma} \label{lem:mono3}
$ (\partial P / \partial y_k +
\partial P / \partial z_k)|_{(w,\yvec,\zvec)} < 0$
if $(\yvec,\zvec) \in V$ and $y_k = z_k$.
\end{lemma}
\begin{proof}
\begin{eqnarray*}
\frac{\partial P}{\partial y_k} +
\frac{\partial P}{\partial z_k}
& = &
\frac{\partial}{\partial y_k} \left[
\prod_{i=1}^n (w+y_i) \right] \; - \;
\sum_{j=2}^n \frac{\partial P_j}{\partial y_k}
\; - \; \frac{\partial P_k}{\partial z_k} \\
& = &
\frac{1}{w+y_k} \prod_{i=1}^n (w+y_i) \; - \;
\frac{1}{y_k} \sum_{j=2}^{k-1} P_j \; - \;
\frac{1}{w+y_k} \sum_{j=k+1}^n P_j \; - \;
\frac{1}{z_k} P_k \\
& < &
\frac{1}{w+y_k} \prod_{i=1}^n (w+y_i) \; - \;
\frac{1}{w+y_k} \sum_{j=2}^{k-1} P_j \; - \;
\frac{1}{w+y_k} \sum_{j=k+1}^n P_j \; - \;
\frac{1}{w+y_k} P_k \\
& = & \frac{P(w,\yvec,\zvec)}{w+y_k}  \\
& = & 0
\end{eqnarray*}
\end{proof}
\begin{cor}
\label{cor:wA2}
$w$ is monotonically increasing in $A_2$.
\end{cor}
\begin{proof}
If we change $A_2$ from $k$ to $k+1$, this
changes $$\yvec = (1,2,\ldots,k-1,k,k,\ldots,k)$$
into $$\yvec' = (1,2,\ldots,k-1,k,k+1,\ldots,k+1)$$
and it changes
$$\zvec = (0,0,\ldots,0,\min(A_1,k),\min(A_1+1,k),
\ldots,k,k)$$ into
$$\zvec' = (0,0,\ldots,0,\min(A_1,k+1),\min(A_1+1,k+1),
\ldots,k+1,k+1).$$  Letting $\mathbf{e}_j^{(y)}$ denote
a unit vector in the $+y_j$ direction, and
$\mathbf{e}_j^{(z)}$ a unit vector in the
$+z_j$ direction, the
direction of change is expressed by the vector
\[
\xvec = (\yvec',\zvec') - (\yvec,\zvec)
= \sum_{k+1 \le j < A_2} \mathbf{e}_j^{(y)}
+ \sum_{\max(k+1,A_2) \le j} (\mathbf{e}_j^{(y)} +
\mathbf{e}_j^{(z)})
\]
and $\partial_{\xvec} P$ is negative, by the preceding two
lemmas.  By Claim~\ref{claim:dirdiv}, this means $w$
increases monotonically as we move along this path.
\end{proof}

%% file: appdx4.tex
\section{Proof of Claim \ref{claim:spenc}}\label{app:prclaim}
To see the finiteness of $X_t$, we need to show that $\sum_{n=1}^\infty o_n=\infty$ a.s. But this follows easily
from the fact that $\sum_{n=1}^\infty n^{-1}=\infty$. To see (\ref{eq:spenc}), we use the following argument,

The standard birth process is equivalent to the following process: Start with one cell at time $0$. At each time, every cell
multiplies with rate $2$. $X_t$ is the number of cells at time $t$.
\begin{lemma}[Joel Spencer]\label{lem:toch}
For every $t>0$ and every positive integer $k$,
$
\E(X_t^k)<\infty.
$
\end{lemma}
\begin{proof}
Let $T=(V,E)$ be an infinite rooted binary tree, and let $\{u_e\}_{e\in E}$ be i.i.d. exponential variables with expected value $0.5$.
Then, $X_t$ is dominated by the size of
\[
Y_t=\left\{v\in V:\ \sum_{e\in\gamma(v)}u_e<t
\right\}
\]
where $\gamma(v)$ is the path from the root to $v$.
For $v$ of depth $n$, the probability that $v$ is in $Y_t$ is
\[
\prob(\mbox{Poisson}(2t)>n)=o\left(\left((n/2)!\right)^{-1}\right)
\]
Therefore,
\[
\prob(|Y_t|>h)\leq\sum_{h=\left[\frac{\log h}{log 2}\right]}^\infty 2^n\prob(\mbox{Poisson}(2t)>n)
\leq C(t)h^{-\log h}
\]
and therefore all moments of $X_t$ are finite.
\end{proof}
Claim \ref{claim:spenc} will follow from Chebyshef if we show that
\begin{equation}\label{eq:et1}
\E(X_{t_2}|X_{t_1})=X_{t_1}O\left(e^{2(t_2-t_1)}\right)
\end{equation}
and
\begin{equation}\label{eq:vt1}
\var(X_{t_2}|X_{t_1})=X_{t_1}O\left(e^{4(t_2-t_1)}\right)
\end{equation}
for $t_2>t_1$.
To show (\ref{eq:et1}) and (\ref{eq:vt1}), it is enough to show that $\E(X_{t})=O(e^{2t})$ and $\var(X_{t})=O(e^{4t})$.
$\E(X_{t})=O(e^{2t})$ follows because $f(t)=\E(X_t)$ satisfies the differential equation
\[
\frac{df}{dt}=2f(t).
\]
$\var(X_{t})=O(e^{4t})$ follows using the exact same martingale argument as in Claim \ref{claim:mati}.

%% file: main.bbl
\begin{thebibliography}{10}

\bibitem{ACL02}
W.~Aiello, F.~Chung, and L.~Lu.
\newblock Random evolution of massive graphs.
\newblock In {\em Handbook of Massive Data Sets}, pages 97--122. Kluwer, 2002.

\bibitem{BA-RMP02}
R.~Albert and A.-L. Barab\'{a}si.
\newblock Statistical mechanics of complex networks.
\newblock {\em Rev. Mod. Phys.}, 74:47--97, 2002.

\bibitem{Aldous-SCN}
D.~J. Aldous.
\newblock A stochastic complex network model.
\newblock {\em Electron. Res. Announc. Amer. Math. Soc.}, 9:152--161, 2003.

\bibitem{BarabasiAlbert99}
A.-L. Barab\'{a}si and R.~Albert.
\newblock Emergence of scaling in random networks.
\newblock {\em Science}, 286:509--512, 1999.

\bibitem{BBBCR03}
N.~Berger, B.~Bollob\'{a}s, C.~Borgs, J.~T. Chayes, and O.~Riordan.
\newblock Degree distribution of the {FKP} network model.
\newblock In {\em International Colloquium on Automata, Languages and
  Programming}, 2003.

\bibitem{BBCR03}
B.~Bollob\'{a}s, C.~Borgs, J.~Chayes, and O.~Riordan.
\newblock Directed scale-free graphs.
\newblock In {\em Proceedings of the 14th ACM-SIAM Symposium on Discrete
  Algorithms}, pages 132--139, 2003.

\bibitem{BolloRior02}
B.~Bollob\'as and O.~Riordan.
\newblock Mathematical results on scale-free random graphs.
\newblock In {\em Handbook of Graphs and Networks}, Berlin, 2002. Wiley-VCH.

\bibitem{BRST01}
B.~Bollob\'{a}s, O.~Riordan, J.~Spencer, and G.~E. Tusnady.
\newblock The degree sequence of a scale-free random graph process.
\newblock {\em Random Structure and Algorithms}, 18:279--290, 2001.

\bibitem{CD-HOT99}
J.~M. Carlson and J.~Doyle.
\newblock Highly optimized tolerance: a mechanism for power laws in designed
  systems.
\newblock {\em Phys. Rev. E}, 60:1412, 1999.

\bibitem{CF01}
C.~Cooper and A.~M. Frieze.
\newblock A general model of web graphs.
\newblock In {\em Proceedings of 9th European Symposium on Algorithms}, pages
  500--511, 2001.


\bibitem{DorogMendes02}
S.~N. Dorogovtsev and J.~F.~F. Mendes.
\newblock Evolution of networks.
\newblock {\em Adv. Phys.}, 51:1079, 2002.

\bibitem{Polya}
F.~Eggenberger and G.~P\'{o}lya.
\newblock \"{U}ber die statistik verketteter.
\newblock {\em Vorg\"{a}nge. {\rm Zeitschrift Agnew. Math. Mech.}}, 3:279--289,
  1923.

\bibitem{FKP02}
A.~Fabrikant, E.~Koutsoupias, and C.H. Papadimitriou.
\newblock Heuristically optimized trade-offs: a new paradigm for power laws in
  the internet.
\newblock In {\em International Colloquium on Automata, Languages and
  Programming}, pages 110--122, 2002.

\bibitem{Falout3}
M.~Faloutsos, P.~Faloutsos, and C.~Faloutsos.
\newblock On the power-law relationships of the {Internet} topology.
\newblock {\em Comput. Commun. Rev.}, 29:251, 1999.

\bibitem{GovinTangmu2000}
R.~Govindan and H.~Tangmunarunkit.
\newblock Heuristics for {Internet} map discovery.
\newblock In {\em Proceedings of INFOCOM}, pages 1371--1380, 2000.

\bibitem{KenyonSchab}
C.~Kenyon and N.~Schabanel.
\newblock Personal communication.

\bibitem{KRRSTU}
R.~Kumar, P.~Raghavan, S.~Rajagopalan, D.~Sivakumar, A.~Tomkins, and E.~Upfal.
\newblock Stochastic models for the web graph.
\newblock In {\em Proc. 41st {IEEE Symp. on Foundations of Computer Science}},
  pages 57--65, 2000.

\bibitem{MEJN-SIAM}
M.~E.~J. Newman.
\newblock The structure and function of complex networks.
\newblock {\em SIAM Review}, 45:167--256, 2003.

\bibitem{Price}
D.~J. de~S.~Price.
\newblock A general theory of bibliometric and other cumulative advantage
  processes.
\newblock {\em J. Amer. Soc. Inform. Sci.}, 27:292--306, 1976.


\bibitem{Simon}
H.~A. Simon.
\newblock On a class of skew distribution functions.
\newblock {\em Biometrika}, 42(3/4):425--440, 1955.

\bibitem{Yule}
G.~U. Yule.
\newblock A mathematical theory of evolution, based on the conclusions of {Dr.
  J. C. Willis}.
\newblock {\em Philos. Trans. Roy. Soc. London}, Ser. B 213:21--87, 1924.

\bibitem{Zipf}
G.~K. Zipf.
\newblock {\em {Human Behavior and the Principle of Least Effort}}.
\newblock Addison-Wesley, Cambridge,MA, 1949.

\vfill
\end{thebibliography}
